\documentstyle[psfig,aps]{revtex}
\begin{document}
\twocolumn
\draft    %JPB

\title{Atom loss from Bose-Einstein condensates due to Feshbach resonance}

\author{V. A. Yurovsky, A. Ben-Reuven,}

\address{School of Chemistry, Tel Aviv University, 69978
Tel Aviv, Israel}

\author{P. S. Julienne and C. J. Williams}

\address{Atomic Physics Division, Stop 8423, National
Institute of Standards and Technology, Gaithersburg, MD 20889,
USA}

\date{\today}
\maketitle    %JPB

\begin{abstract} In recent experiments  on Na Bose-Einstein condensates
[S. Inouye {\it et al}, Nature {\bf 392}, 151 (1998);
J. Stenger {\it et al}, Phys. Rev. Lett. {\bf 82}, 2422 (1999)],
large loss rates were observed when a time-varying magnetic field was
used to tune a molecular Feshbach resonance state near the state of pairs 
of atoms belonging to the condensate many-body wavefunction.
A mechanism is offered here to account for the observed losses, based
on the deactivation of the resonant molecular state by interaction with a
third condensate atom, with a deactivation rate coefficient of magnitude
$\sim 10^{-10}$cm$^{3}/$s.

\end{abstract}
\pacs{03.75.Fi, 32.80.Pj, 32.60.+i, 34.50.Ez, 34.90.+q}

%\narrowtext

Experiments have been carried out recently~\cite{IASMSK98,SIAMSK99}
(see also the review~\cite{TTHK99}) in order to control the interatomic
interaction underlying the properties of a Bose-Einstein condensate (BEC).
One way to achieve this ~\cite{IASMSK98,SIAMSK99} is by tuning a
magnetic field $B$ to modify the two-atom scattering length as predicted
for a $B$-dependent Feshbach resonance~\cite{Verhaar,MVA95}.
The experiments carried out on a Na optical trap measured two distinct
features: (a) the change in scattering length and large collisional atom
losses with a slow sweep of $B$ that stopped short of the resonant field $B_0$,
and (b) a near-catastrophic loss of atom density with a fast sweep of $B$
through the $B_0$ region.
Two groups of investigators~\cite{MJT99,VerhaarLect} have recently proposed
a unimolecular mechanism to explain the latter feature as due to a fast
sweep-induced transfer of population from the ground trap state of an atom
pair to excited states.

We present here another possible mechanism of atom loss, which
is important when the field is changed slowly, and may play a role
in the fast-sweep case as well. The temporary occupation of the
vibrationally-excited molecular state $m$, coupled by the resonance to
atom pairs in the condensate state $a$, is converted to a stable molecular
dimer state $d$ by a deactivating inelastic collision with a third atom:
\begin{eqnarray}
&&\text{Na}+\text{Na}\rightleftharpoons\text{ Na}_{2}\left( m\right)
\label{RCol}
\\
&&\text{Na}_{2}\left( m\right) +\text{Na}\rightarrow
\text{Na}_{2}\left( d\right) +\text{Na}+\Delta E \label{SCol}
\end{eqnarray}
where $\Delta E$ is the excess kinetic energy released in the inelastic
process. The deactivation states can be lower-lying rovibrational levels
in the same spin state as the resonant level, or levels belonging to
another spin state.  While step (1), standing alone, is completely
reversible, step (2) is irreversible, owing to the rapid ejection of
the products from the condensate.
The latter step usually provides sufficient kinetic energy to freely
escape the trap, and is characterized by a deactivation rate coefficient
$2\gamma $ (in units of cm$^3/$s). The two steps described above
may be followed by a third one --- secondary collisions of the
products with condensate atoms.

Except for $\gamma $, the magnitude of which can be extracted from
fitting the experimental data, all expressions below can be derived
analytically, using previously-determined parameters.  A variational method
was used to derive a set of three coupled Gross-Pitaevskii equations for
the atomic condensate state $\varphi _{a}\left( {\bf r},t\right)$,
the resonant molecular state $\varphi _{m}\left( {\bf r},t\right)$,
and a representative deactivation state $d$. The latter can be decoupled out,
leaving terms dependent on $\gamma$ in the remaining two equations
(similar to those recently used
by Timmermans {\it et al.}~\cite{TTHK99,TTCHK98}):
\begin{eqnarray}
i\hbar \dot{\varphi }_{a}&&=\left( {1\over 2m}\hat{{\bf p}}^{2}
+V_{a}\left( {\bf r}\right) +\mu _{a}B\left( t\right) +
U_{a} |\varphi_{a}|^{2}\right) \varphi _{a}
\nonumber
\\
&&
+U_{am}|\varphi _{m}|^{2} \varphi _{a}
+2g^{*}\varphi^{*}_{a}\varphi_{m}
-i\hbar \gamma |\varphi_{m}|^{2}\varphi_{a}
\label{GPa}
\\
i\hbar \dot{\varphi }_{m}&&=\left( {1\over 4m}\hat{{\bf p}}^{2}
+V_{m}\left( {\bf r}\right) + \mu _{m}B\left( t\right)
+U_{m}|\varphi _{m}|^{2}\right)\varphi_{m}
\nonumber
\\
&&
+U_{am}|\varphi_{a}|^{2}
\varphi_{m}+g\varphi^{2}_{a}
-i\hbar \gamma |\varphi _{a}|^{2}\varphi_{m} \label{GPm}
\end{eqnarray}
provided the deactivation state $d$ never accrues a significant population.\
Here $m$ is the mass of the Na atom; $V_{a}\left( {\bf r}\right)$
and $V_{m}\left( {\bf r}\right)$ are, respectively, the atomic and molecular
optical trapping potentials ($V_m$ including the resonance
detuning for zero magnetic field); ${\bf r}$ is the position in the trap;
$B\left( t\right)$ is the applied homogeneous magnetic field;
$\mu _{a}$  and $\mu _{m}$  are, respectively, the atomic and molecular
magnetic moments; and $U_{a}$, $U_{m}$, and $U_{am}$ are, respectively,
zero-momentum atom-atom, molecule-molecule and atom-molecule interactions,
proportional to the corresponding elastic scattering lengths.
The coupling constant $g$ responsible for the atom-Feshbach coupling
is closely related to the parameter $\Delta$~\cite{IASMSK98,SIAMSK99}
that characterizes the strength of the resonance shape as a function of
the field B:
\begin{equation}
|g|^{2}=2\pi \hbar ^{2}|a_{a}|\mu \Delta /m
\end{equation}
where $a_{a}$ is the off-resonance triplet scattering length.
Calculations~\cite{MJT99,AV99} give $a_a$ = 3.4nm, $\mu = \mu_m-2\mu_a
= 3.3\mu_B$ (where $\mu_B = 9.27\times 10^{-24}$J/T is the Bohr
magneton), and $\Delta =$ 0.001mT and 0.1mT, respectively, 
for the two resonances observed~\cite{IASMSK98,SIAMSK99}
at 85.3mT (853G) and 90.7mT (907G). These $\Delta$ values agree with
the value measured for the 90.7mT resonance, and with the
indirectly inferred order of magnitude for the 85.3mT resonance.

The analysis below neglects the kinetic energy terms in Eqs.\
(\ref{GPa}) and (\ref{GPm}), in accord with the Thomas-Fermi approximation.
This reduces (\ref{GPa}) and (\ref{GPm}) to a system of
ordinary differential equations in $\varphi _{a}\left( {\bf r},t\right) $
and $\varphi _{m}\left( {\bf r},t\right) $ that depend parametrically
on ${\bf r}$. The resulting set of four real equations
can then be solved numerically, for a given value of $\gamma $,
starting with a Thomas-Fermi
distribution or with a homogeneous (${\bf r}$-independent) distribution
equal to the mean trap density.

We have carried out such numerical solutions for both types of MIT
experiments \cite{IASMSK98,SIAMSK99}, using the homogeneous initial
conditions. In the slow-sweep experiment the time-dependent magnetic
field was linearly changed from an initial value of $B$ to
a final value closer to resonance, and then the density was
measured by shutting off the trap field and letting the
condensate expand. Repeating the experiment at various final
values of the magnetic field, the loss was plotted as a function
of this final value. It was noted \cite{SIAMSK99} that the results
fit to a 3-body rate equation for the atomic density
$n\left( {\bf r},t\right) =|\varphi _{a}\left( {\bf r},t\right) |^{2}$,
of the type $\dot{n}\approx -K_{3}n^{3}$.

Analysis of Eqs.\ (\ref{GPa}, \ref{GPm}) shows that, out of the four real
coupled equations for $\varphi_{a}$ and $\varphi_{m}$, a single rate
equation for the atomic density
$n\left( {\bf r}, t\right) =|\varphi _{a}\left( {\bf r},t\right) |^{2}$
can be extracted, if the following
condition (called here the ``fast decay'' approximation)
\begin{equation}
|g|^{2}\ll \hbar ^{2}\gamma ^{2} n\left( {\bf r}, t\right) \label{FDA}
\end{equation}
(i.e., $n_{m}\left( {\bf r},t\right) =|\varphi _{m}\left( {\bf r},t\right)
|^{2}\ll n\left( {\bf r},t\right) $) is observed.
The resulting rate equation attains the nonlinear form
$\dot{n}\left( {\bf r},t\right) =-F\left( n\left( {\bf r},t\right),
{\bf r},t\right)$,
where
\begin{equation}
F\left( n,{\bf r},t\right) ={6|g|^{2}\gamma n{ } ^{3}\over
\left\lbrack V\left( {\bf r}\right) -\mu B\left( t\right)
\right\rbrack ^{2}+\left( \hbar \gamma n\right) { } ^{2}}
\label{RE}
\end{equation}
and $V\left( {\bf r}\right) = 2V_{a}\left( {\bf r}\right)
- V_{m}\left( {\bf r}\right)$.
Eq.\ (\ref{RE}) can be expressed in terms of a collision cross section of
the Breit-Wigner shape (dependent on the position ${\bf r}$ in the
trap) for zero-momentum collisions, in which $2\gamma n$ is the width of
the output channel, and the width of the input channel is proportional
to $|g|^{2}$.
This observation establishes a link between the macroscopic approach used here
and microscopic approaches, treating the loss rate as a collision process.

It follows from Eq.\ (\ref{RE}), that very close to resonance
(where $B\left( t\right) $ is within $1\mu $T of a resonance), the
behavior of Eq.\ (\ref{RE}) changes its form gradually from that of a
3-body to that of an effective 1-body rate equation.
However, all values of $B$ used for the slow-sweep experiments of the
MIT group lie outside this narrow region, where the
``off-resonance'' condition
\begin{equation}
\hbar \gamma n\left( {\bf r}, t\right)
\ll |V\left( {\bf r}\right) -\mu B|\qquad \label{NRA}
\end{equation}
holds. Given this condition, one can write Eq.\ (\ref{RE})
(to a very good approximation) in the 3-body form
$F\left( n,{\bf r},t\right) =K_{3}\left( {\bf r},t\right) n^{3}$
with
\begin{equation}
K_{3}\left( {\bf r},t\right) ={12\pi \hbar ^{2}|a_{a}|\gamma \Delta
\over m\mu \left\lbrack B\left( t\right) -B_{0}\left( {\bf r}\right)
\right\rbrack { } ^{2}} \label{K3}
\end{equation}
where $B_{0}({\bf r})$ is the local resonance value of $B$.
We shall refer to the combination of Eqs.\ (\ref{FDA}) and
(\ref{NRA}), leading to Eq.\ (\ref{K3}), as the ``3-body''
approximation. This expression is similar to the one derived earlier
\cite{TTHK99,TTCHK98} up to a factor $3/2$, which reflects the fact that
3 atoms are lost per one deactivating collision.

For $K_{3}$  given by Eq.\ (\ref{K3}), the rate equation may be
solved analytically in the form
\begin{eqnarray}
n\left( {\bf r},t\right) =n\left( {\bf r},t_{0}\right) \biggl\lbrack
1+24\pi \hbar ^{2}|a_{a}|\Delta \gamma n^{2}\left( {\bf r},t_{0}\right)
\nonumber
\\
\times \left( t-t_{0}\right) /\left( m\mu \dot{B}^{2}t t_{0}\right)
\biggr\rbrack ^{-1/2} \label{nt}
\end{eqnarray}
where $\dot{B}$ is the magnetic-field ramp speed,
and the starting time $t_{0}$ is measured from
the extrapolated resonance-crossing time as $t=0$.
The graphs shown in
Figs.\ 1 and 2 pertain to the slow-sweep experiment conducted at MIT,
and more particularly to the strong 90.7 mT resonance. This
resonance has been approached from below with two ramp speeds, and
from above --- with one. The difference between Eqs.\
(\ref{K3},\ref{nt}) and the results of a direct numerical solution of
Eqs.\ (\ref{GPa}) and (\ref{GPm}) for all ramp speeds in the $K_{3}$ plots
(Fig.\ 1) is so small, that the corresponding plots are indistinguishable.
The remaining atomic density, at the moment $t$ of
stopping the ramp (see Fig.\ 2), was also calculated, using the
homogeneous-density initial condition,
starting from a $B$ value of 89.4mT on approach from below and 91.6mT
from above, for which the initial mean densities were extracted from the
experimental data \cite{SIAMSK99}. The graphs clearly show a best fit
with $\gamma $ of the order of $10^{-10}$cm$^{3}/$s, which (given a density
of about $10^{15}$cm$^{-3}$) implies a deactivation time of $\sim 10^{-5}$s.

The magnitude of the inelastic rate coefficient $2\gamma$
needed to explain the data is quite plausible.
It is two orders of magnitude smaller than the upper limit
allowed by the unitarity constraint on the $S$-matrix.
According to this condition \cite{MJ89}, in the limit of
small momentum, $2\gamma \le \hbar \lambda /m$,
where $\lambda $ (the de Broglie wavelength) is in turn limited by
the trap dimensions. This requirement puts an upper bound of
$2.5\times 10^{-8}$cm$^{3}/$s to $2\gamma$ in the case considered here.
Furthermore, our estimate of $10^{-10}$cm$^{3}/$s
for $\gamma$ is consistent with calculations carried out recently
~\cite{Forrey99} on the rate of vibrational deactivation in
H$_{2}+$He ultracold collisions, producing a comparable result
in spite of the large energy gaps in H$_{2}$.

It is interesting to compare the outcome of the deactivation
mechanism described here with that of other loss mechanisms based on a
unimolecular process (not involving a third atom)~\cite{MJT99,VerhaarLect},
proposed for the second type of MIT experiment with the fast sweep of B
through resonance. To see this one can remove the imaginary
term in Eq.\ (\ref{GPa}), and replace $\gamma n$ in
Eq.\ (\ref{GPm}) with an $n$-independent $\Gamma $. In the fast-decay
limit, in which $n_{m}\ll n$, both mechanisms would yield similar
results for $K_{3}$ and $n$ of the slow-sweep experiments, if similar decay
times ($\sim 10^{-5}$s) were to be reckoned in both processes.

The 3-body approximation does not hold very close to resonance,
and is therefore inapplicable to a description of the fast-sweep
experiment, in which the Zeeman shift was swept rapidly through the
resonance, causing the most dramatic loss rates (see Refs.\
\cite{IASMSK98,SIAMSK99}). Nevertheless, maintaining only the fast
decay approximation, Eq.\ (\ref{FDA}), a simple analytical expression
can be obtained also for this kind of measurement. Assuming the
magnetic field variation is long enough to allow taking the asymptotic
time limits, reached under the condition
$\mu \delta B\gg \hbar \gamma n$
where $\delta B$ is the range of variation of $B$ on either side of the
resonance, the ramp starting and stopping times can be extended to
$\pm \infty $.
One then obtains at all positions ${\bf r}$
\begin{equation}
n\left( {\bf r},\infty \right) ={n\left( {\bf r},-\infty \right)
\over 1+s n\left( {\bf r},-\infty \right) },\quad s={12\pi ^{2}
\hbar|a_{a}|\over m}{\Delta \over |\dot{B}|} . \label{n}
\end{equation}
This asymptotic result (as opposed to the one mentioned earlier with
regard to the slow-sweep MIT experiment) is independent of $\gamma $.

Assuming, as before, a homogeneous initial density within the trap,
Eq.\ (\ref{n}) also describes the loss of the total population
$N\left( t\right) = \int n\left( {\bf r},t\right) d^{3}r$.
An analytical asymptotic expression for the total population
can also be found when the homogeneous distribution is replaced
by the Thomas-Fermi one (given in Ref.\ \cite{MADKDK96}), in which case
\begin{eqnarray}
{N\left( \infty \right) \over N_{0}\left( -\infty \right) }={15\over
2sn{ } _{0}}\biggl\lbrack {1\over 3}+{1\over sn{ } _{0}}-{1\over
2sn{ } _{0}}\sqrt{1+{1\over sn{ } _{0}}} \nonumber
\\
\times \ln \bigglb(\left( \sqrt{1+{1\over sn{ } _{0}}}+1\right)
/\left( \sqrt{1+{1\over sn{ } _{0}}}-1\right) \biggrb)\biggr\rbrack
\label{gd}
\end{eqnarray}
where $n_{0}$  is the maximum initial density in the center of the trap.

The analytical results of Eq.\ (\ref{n}), together with the direct
numerical solutions of Eqs.\ (\ref{GPa}) and (\ref{GPm}), are compared
in Fig.\ 4 with the results of the fast-sweep experiment
\cite{IASMSK98,SIAMSK99}. The numerical results show that if the
deactivation time is not small enough compared to the sweep time, and the
magnetic field variation is not large enough, the assumptions underlying
Eqs.\ (\ref{n}) and (\ref{gd})
do not hold, and thus the condensate loss will depend on $\gamma $.
These calculations show a rough agreement with the
measured loss rates for values of $\gamma $ of the orders of
$10^{-9}$-$10^{-11}$cm$^{3}/$s, though these results are not as sensitive
to $\gamma $ as were the fits to the slow-sweep experiments discussed
above.

The deactivation reaction Eq.\ (\ref{SCol}) produces $4\gamma nn_{m}$
``hot'' particles (atoms and molecules) in a unit volume per unit time.
Traversing a distance $b$ at a speed $v$, these particles create new hot
particles in a cascading process. The density of hot particles
may then be estimated as
$n_{h}\approx 4\gamma nn_{m}b/v \exp\left( b\sigma \left( n+n_{m}\right)
\right)$, where $\sigma $ is the elastic collision cross-section,
considered here identical for atom-atom and atom-molecule collisions.
The rate of additional loss due to these secondary collisions is
estimated as $v\sigma nn_{h}$ for the atomic condensate
and $v\sigma n_{m}n_{h}$ for the molecular condensate. This loss rate
depends essentially on the product $\sigma b$. 
A conservative estimate of this product is obtained by taking
$\sigma = 0.5\times 10^{-12}$cm$^{2}$
(the elastic Na-Na cross section at 0.1K \cite{CD94}, a typical vibrational
deactivation energy), and $b=10^{-3}$cm (the condensate-cloud radius).
A more realistic value of this product may be higher on account of
the larger axial condensate length, the rise of the elastic cross section
at energies below $\sim 10$mK~\cite{CD94}, and a possible contribution
of inelastic collisions. Two values of $\sigma b$ are therefore used in
Fig.\ 4. This figure shows that the effect of secondary collisions, 
calculated for a homogeneous density, may appreciably increase the
condensate loss, while the use of a Thomas-Fermi density
[see Eq.\ (\ref{gd})] should decrease it.

The authors are most grateful to Wolfgang Ketterle and Joern Stenger
for helpful information regarding the MIT experiments.

\begin{figure}

\vspace{1.5cm}

\psfig{figure=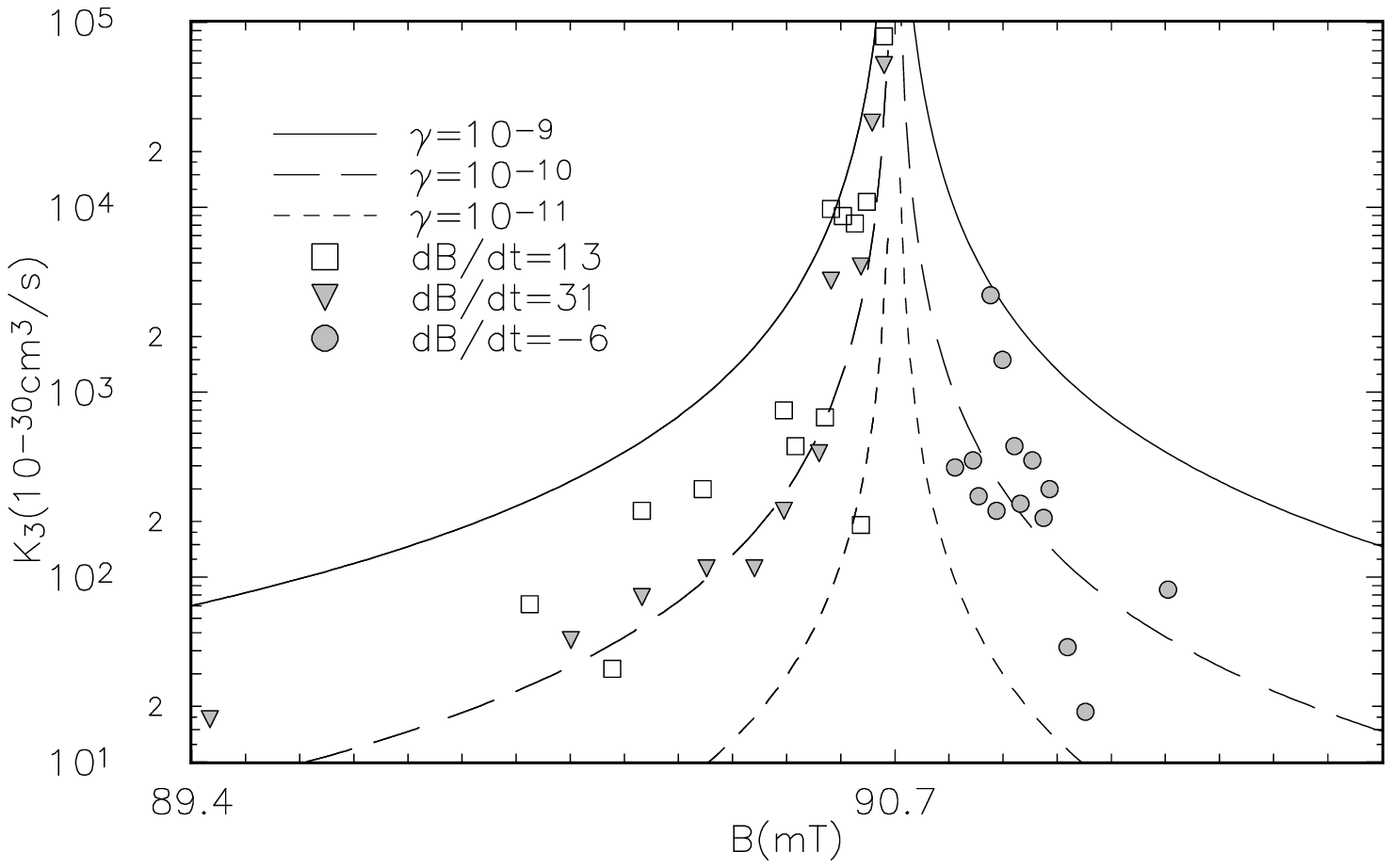,width=3.375in}
\caption{The 3-body rate coefficient ($K_{3}$) vs. the stopping
value of the magnetic field, calculated  with 3 different values of
the deactivation rate $\gamma $ (in units of cm$^{3}$/s), on approaching
the resonance from below or above. These are compared with the experimental
results (Ref.\ \protect\cite{SIAMSK99}). Ramp speeds shown are in units
of mT/s.}

\bigskip

\psfig{figure=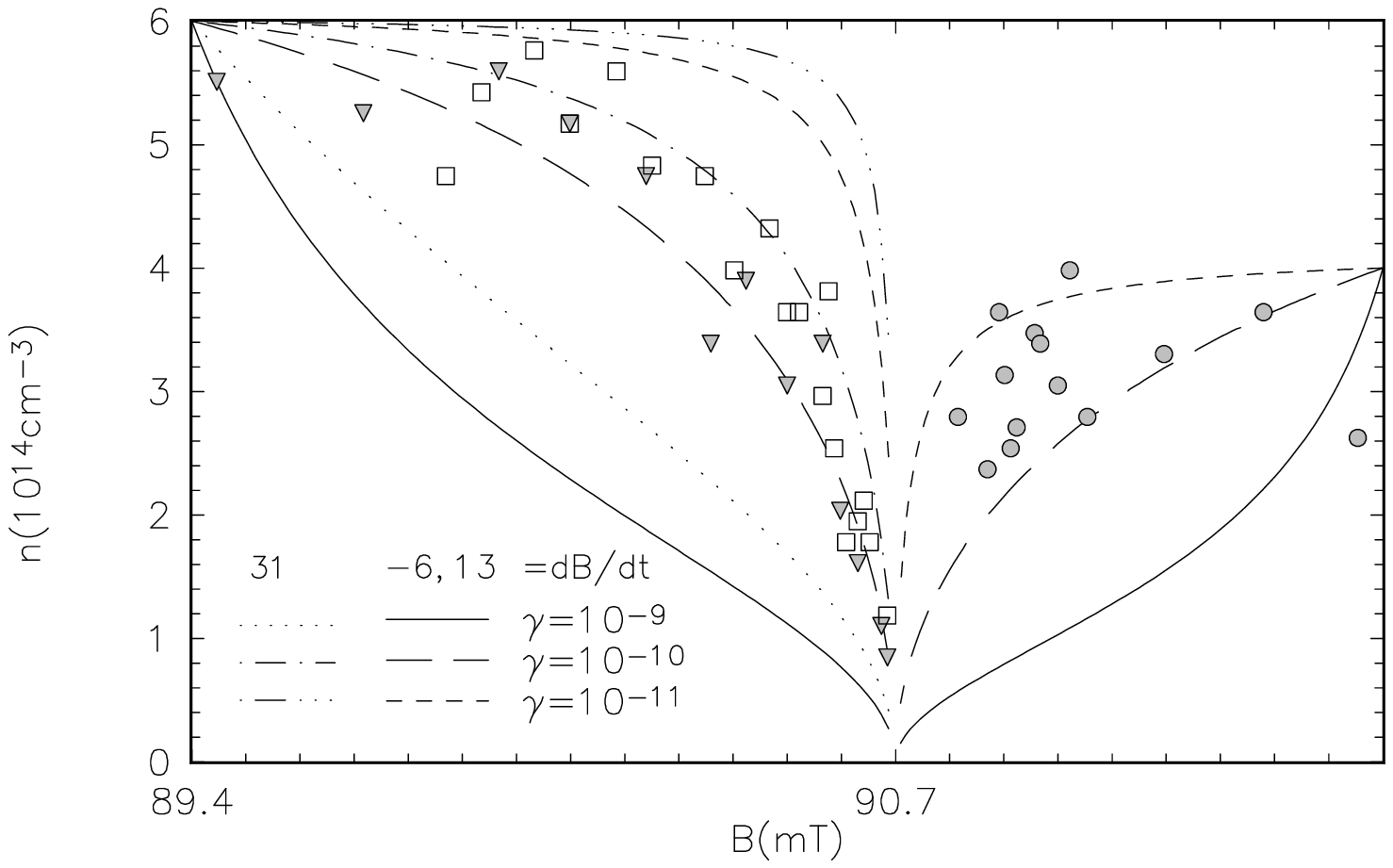,width=3.375in}
\caption{The surviving mean density vs. the stopping value of
the magnetic field, calculated  with 3 different values of the ramp
speed $\dot{B}$ (in mT/s). Other notations as in Fig.\ 1.}
\newpage

\psfig{figure=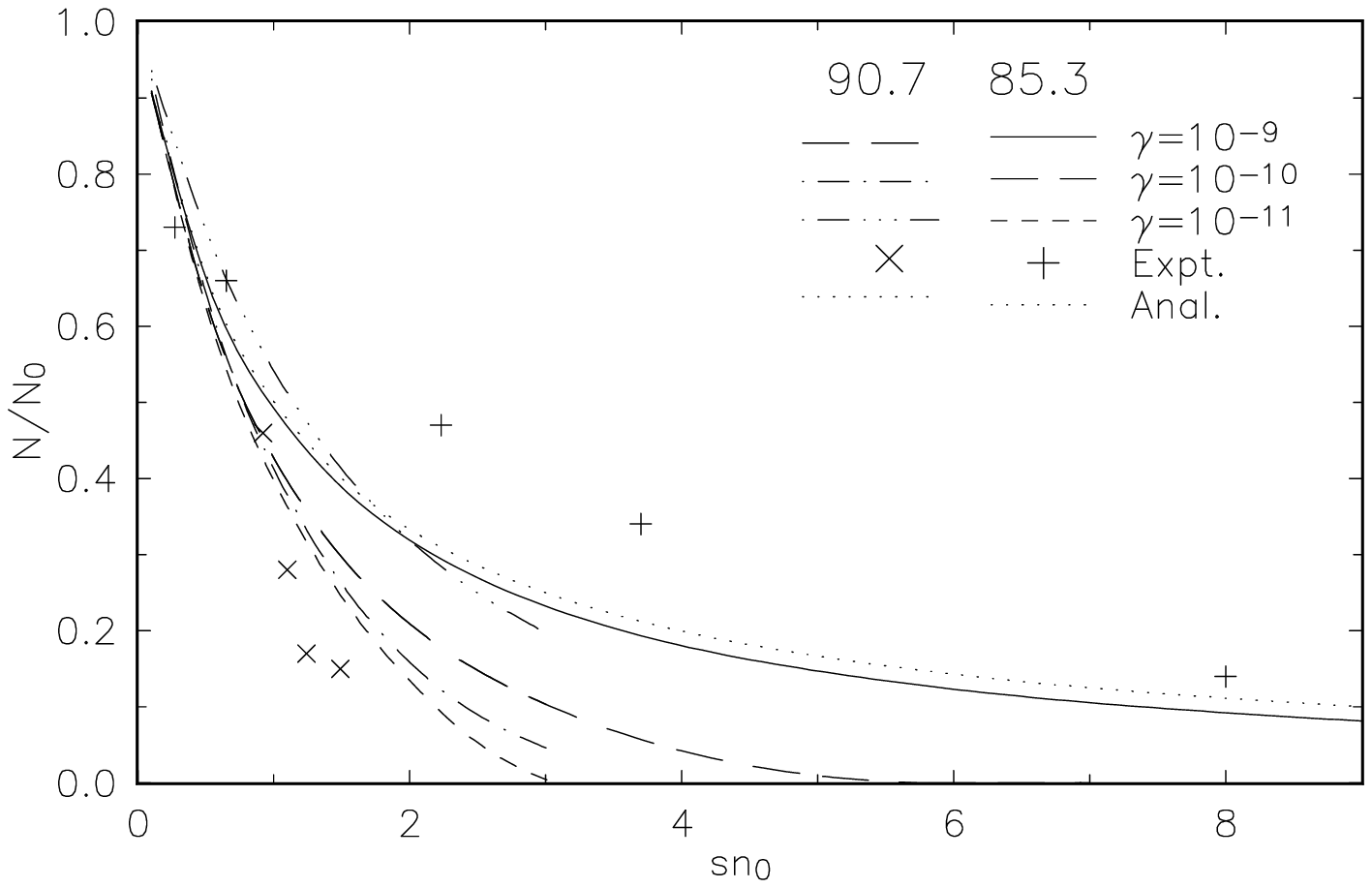,width=3.375in}
\caption{Ratio of surviving trap population $N$ to the initial one
$N_{0}$ for
the 85.3mT (853G) and 90.7mT (907G) resonances in the homogeneous-density
approximation vs. $s n_{0}$ (where the parameter $s$ is defined by Eq.\
(\protect\ref{n}) and $n_{0}$ is the initial density). The curves show
calculations carried out for different magnitudes of the coefficient
$\gamma $ (in units of cm$^{3}$/s), and with the analytical result of the
fast-decay approximation Eq.\ (\protect\ref{n}) (dots). The plots for the
90.7mT resonance with $\gamma =10^{-9}$cm$^{3}/$s and for the 85.3mT
resonance with $\gamma =10^{-10}$cm$^{3}/$s are practically
indistinguishable. The calculations are compared
with the results of the MIT fast-sweep experiment (Ref.\
\protect\cite{SIAMSK99}).}

\bigskip

\psfig{figure=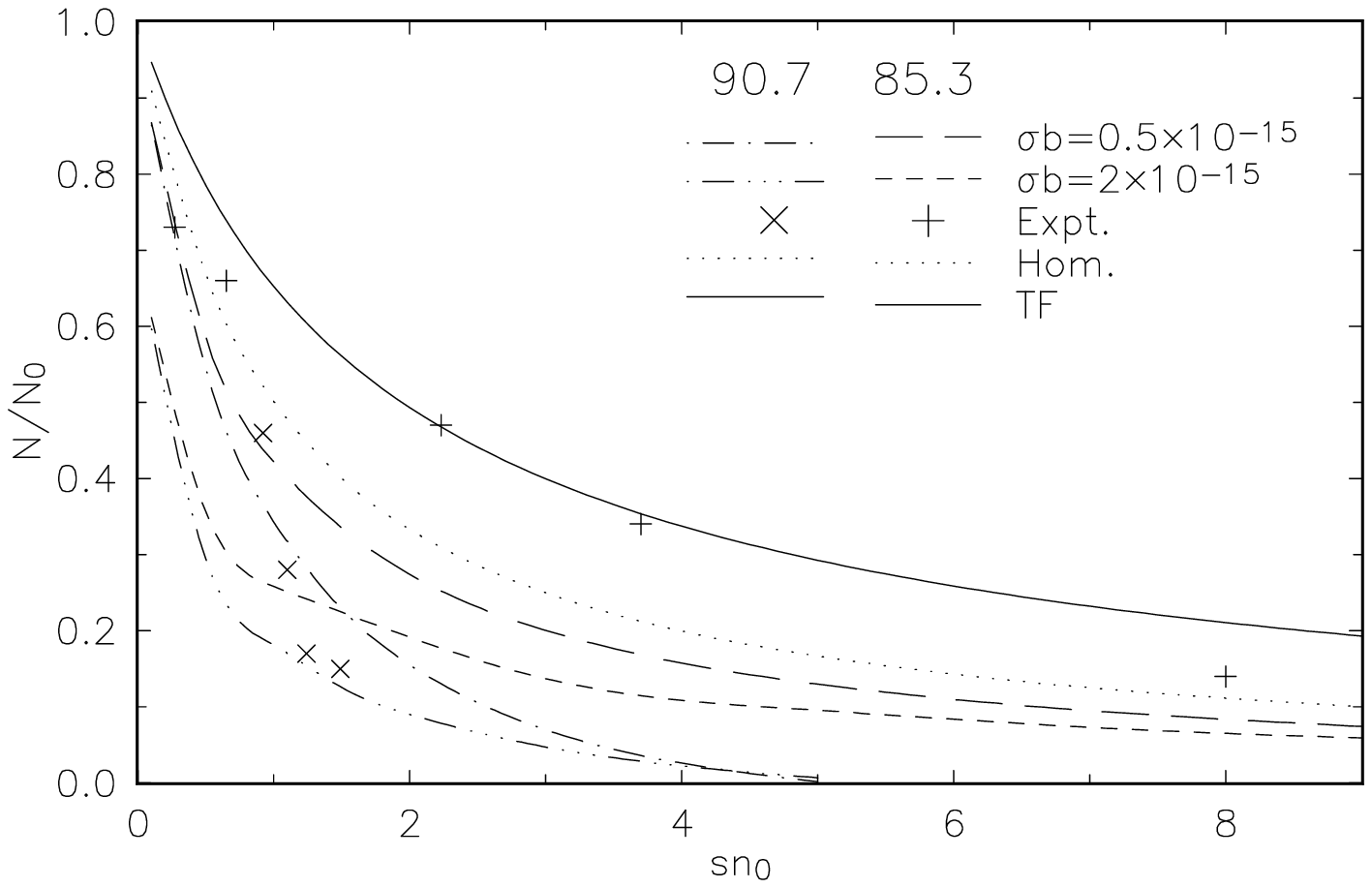,width=3.375in}
\caption{Same as Fig.\ 3, but taking account of secondary collisions
for $\gamma =10^{-9}$cm$^{3}/$s and various values of the parameter
$\sigma b$ (in units of cm$^{3}$).
The analytical results (without taking account of secondary collisions)
for the Thomas-Fermi distribution Eq.\
(\protect\ref{gd}) (solid line) and the homogeneous distribution
 Eq.\ (\protect\ref{n}) (dots) are plotted as well.}

\end{figure}

\begin{references}
%\Bibliography{99}   %JPB

\bibitem{IASMSK98}S. Inouye, M. R. Andrews, J. Stenger, H.-J.
Miesner, D. M. Stamper-Kurn, and W. Ketterle, Nature {\bf 392}, 151
(1998).

\bibitem{SIAMSK99}J. Stenger, S. Inouye, M. R. Andrews, H.-J.
Miesner, D. M. Stamper-Kurn, and W. Ketterle, Phys. Rev. Lett. {\bf 82},
2422 (1999).

\bibitem{TTHK99}E. Timmermans, P. Tommasini,M. Hussein, and A.
Kerman, Phys. Rep., to be published (1999).

\bibitem{Verhaar}E. Tiesinga, B.J. Verhaar, and H.T.C.
Stoof,Phys. Rev. {\bf A47}, 4114 (1993); E. Tiesinga, A.J. Moerdijk,
B.J. Verhaar, and H.T.C. Stoof, Phys. Rev. {\bf A46}, R1167 (1992).

\bibitem{MVA95}A. J. Moerdijk, B. J. Verhaar, and A. Axelsson,
Phys.\ Rev.\ A {\bf 51}, 4852 (1995).

\bibitem{MJT99}F. H. Mies, P. S. Julienne, and E. Tiesinga,
prepublication manuscript (1999).

\bibitem{VerhaarLect}F. A. van Abeelen, and B. J. Verhaar, 
"Giant losses in a recent Feshbach resonance scattering experiment at MIT",
Workshop on Formation of Cold Molecules, $1-5$ March 1999, Les Houches,
France (unpublished).

\bibitem{TTCHK98}E. Timmermans, P. Tommasini, R.
C${\rm\hat{o}}$t${\rm\acute{e}}$, M. Hussein, and A. Kerman, Phys.
Rev. Lett. (submitted); cond-mat/9805323.

\bibitem{AV99}F. A. van Abeelen, and B. J. Verhaar, as discussed in
Ref.\ \cite{SIAMSK99}.

\bibitem{MJ89}P. S. Julienne and F. H. Mies, J. Opt. Soc. Am.
B {\bf 6}, 2257 (1989); P. S. Julienne, A. M. Smith, and K. Burnett,
Adv. At. Mol. Opt. Phys. {\bf 30}, 141 (1993).

\bibitem{Forrey99}R. C. Forrey, V. Kharchenko, N. Balakrishnan
and A. Dalgarno, Phys. Rev. A {\bf 59}, 2146 (1999).

\bibitem{MADKDK96}M.-O. Mewes, M. R. Andrews, N. J. van Druten,
D. M. Kurn, D. S. Durfee, and W. Ketterle, Phys. Rev. Lett. {\bf 77},
416 (1996).

\bibitem{CD94}R. C${\rm\hat{o}}$t${\rm\acute{e}}$, and A.
Dalgarno, Phys. Rev. {\bf A50}, 4827 (1994).

\end{references}
\end{document}